\documentclass[12pt,a4paper,final]{iopart}

\usepackage{iopams}
\usepackage{cite}

\expandafter\let\csname equation*\endcsname\relax
\expandafter\let\csname endequation*\endcsname\relax

\usepackage[T1]{fontenc}
\usepackage{amsmath}
\usepackage{iopams}
\usepackage{graphicx}
\usepackage{stackrel}
\usepackage{mathrsfs}
\usepackage[breaklinks=true,colorlinks=true,linkcolor=blue,
urlcolor=blue,citecolor=blue]{hyperref}

\begin{document}

\title{Non-Gaussianity in stochastic transport: phenomenology and modelling}

\author{Ralf Metzler}
\address{Institute of Physics and Astronomy, University of Potsdam, 14476
Potsdam-Golm, Germany}
\ead{rmetzler@uni-potsdam.de}
\author{Aleksei V. Chechkin}
\address{Institute of Physics and Astronomy, University of Potsdam, 14476
Potsdam-Golm, Germany}
\address{Faculty of Pure and Applied Mathematica, Hugo Steinhaus Center,
Wroc{\l}aw University of Science and Technology, Wyspianskiego 27, 50-370
Wroc{\l}aw, Poland}
\address{Akhiezer Institute for Theoretical Physics, 61108 Kharkov, Ukraine}
\ead{chechkin@uni-potsdam.de}

\begin{abstract}
Non-Gaussian shapes, despite a linear form of the mean-squared displacement,
have been observed for the displacement distribution in a large range of
diffusive systems. Stochastic models for such "Brownian yet non-Gaussian"
diffusion will be introduced and discussed. Systems with non-Gaussian,
anomalous diffusion will also be addressed.
\end{abstract}

\section{Introduction}

In their foreword to volume 5 of the famed Theoretical Physics textbook
series, Landau and Lifshitz address the perception of Statistical Physics
\cite{landau}: "Among physicists there exists the widely prevalent fallacy
that Statistical Physics was the least well-founded discipline of Theoretical
Physics. In doing so it is typically referred to the lack of rigorous
mathematical proof of some conclusions in Statistics; one forgets that also
other disciplines of Theoretical Physics contain inexact proofs, but on no
account this is regarded as a signature of an insufficient validity of these
disciplines."

While this perception has not fully vanished (and is partially shared by our
undergraduate students) over the 80 years since this was written\footnote{To
some extent, one might argue that the perception of Statistical Physics is
often (too) closely connected with that of classical Thermodynamics, that in
its axiomatic nature is indeed quite different from other fields of physics.}
Statistical Physics has become a success story, being continuously developed
further, and it has pervaded an ever increasing range of fields, including
the physics of condensed, soft, and biological matter, geophysics, economy,
ecology, or epidemiology. An oustanding acknowledgement of the role of
Statistical Physics has been the 2021 Physics Nobel Prize to Giorgio Parisi
for his work on the Statistical Physics of complex systems.

An avid advocate for Statistical Physics has always been our dear colleague
Paolo Grigolini, whose work we honour in this book. Paolo's work has changed
many aspects of how we understand complex systems, and his most outstanding
contributions are in the Statistical Physics of Non-equilibrium Systems. To
quote from the foreword to volume 10 of the Landau-Lifshitz series, Physical
Kinetics by Lifshitz and Pitaevski in 1978 \cite{landau1}: "In contrast to
the properties of statistical equilibrium systems, kinetic properties are
far more connected with the microscopic interactions of specific physical
objects. This is the cause for the massive diversity of these properties and
the substantially higher level of difficulty of their theoretical analysis."
Indeed, this diversity has been a driving force for exploration for many of
us.

In the field of stochastic processes an outstanding scientist was Elliot
Montroll,
who came to fame when he successfully applied random walk theory to the
behaviour of neutrons in the chain reaction in the Manhattan project. Montroll
was central in formulating continuous time random walks and, ultimately,
setting the scene for
anomalous diffusion \cite{montroll,montroll1,montroll2,scher}. His heritage,
continued by his collaborator Harvey
Scher, his PhD student Mike Shlesinger, and his postdoc Bruce West, is still
central for new developments and applications of stochastic processes. While
the by-now classical textbooks of van Kampen \cite{vankampen} or Gardiner
\cite{gardiner} concentrate on what
we may call "classical dynamics" (Brownian motion, master equations, etc.),
more recent textbooks such as those by Hughes \cite{hughes}, Klafter and Sokolov
\cite{klaso}, or by West, Bologna, and Grigolini \cite{webogri} make strong
cases that anomalous diffusion has become "normal", as stated by Katja
Lindenberg and colleagues \cite{katja}.

In fact, statistical physics is an open, emerging field. One example is the
recent development of stochastic thermodynamics \cite{udo} or efforts to come
up with formulations for non-extensive systems, e.g., systems with long-range
interactions for which the Gibbsian idea of subsystems no longer holds. One
famed mathematical formulation taking such non-extensivity into account was
the generalised entropy formulated by Tsallis \cite{tsallis}. Another field
to mention are ongoing questions on the precise formulation of even classical
entropy, as those posed by Stratonovich \cite{stratonovich}.

A number of concrete and experimentally relevant questions, on which Paolo
has been working, include anomalous diffusion and its origins, the role of
ergodicity, and ageing/non-stationarity in complex systems. In this spirit
we will here address a main field of recent interest, prompted by
the ongoing discovery of non-Gaussian statistics in relatively simple
systems.

In the following we will use the term anomalous diffusion in the sense that
the associated mean squared displacement (MSD) of an ensemble is of
the power-law form\footnote{We will mostly use a one-dimensional notation,
generalisations to higher dimensions are fairly straightforward.}
\begin{equation}
\label{msd}
\langle x^2(t)\rangle=2K_{\alpha}t^{\alpha},
\end{equation}
where $K_{\alpha}$ is the generalised diffusion coefficient of physical
dimension $\mathrm{cm}^2/\mathrm{sec}^{\alpha}$. Depending on the precise
value of the anomalous diffusion exponent $\alpha$, we typically distinguish
subdiffusion ($0<\alpha<1$) and superdiffusion ($\alpha>1$) \cite{report}.
Specific cases are normal ("Brownian" or "Fickian") diffusion for $\alpha=1$
and ballistic motion for $\alpha=2$. The case $\alpha=3$ is that of
Richardson relative diffusion in turbulence \cite{richardson}.

When the increments of a stochastic process are independent and identically
distributed (iid) variables $x_i$ with a finite variance, as originally
considered by Einstein \cite{einstein} and Smoluchowski \cite{smoluchowski},
their normalised sample average $\sqrt{n}[\sum x_i/n-\mu]$ with mean $\mu$
for $n$ approaching infinity, converges in distribution to the normal or
Gaussian probability density function (PDF) \cite{levy,levy1,gnedenko}
\begin{equation}
\label{gauss}
P(x,t)=\frac{1}{\sqrt{4\pi K_1t}}\exp\left(-\frac{x^2}{4K_1t}\right),
\end{equation}
as written here for the case of normal Brownian diffusion with diffusivity
$K_1$. Mathematically, this is due to the law or large numbers or, more
stringently, the central limit theorem \cite{gnedenko}. Violations of one or
more of these conditions lead to anomalous diffusion and non-Gaussian
statistics. Leaving the basin of attraction of the central limit theorem
\cite{bouchaud} causes the loss of the universality of the Gaussian law,
in the sense that details of the considered process become more relevant
and lead to a variety of emerging dynamics \cite{pccp}. We could also say,
things are getting interesting.

\section{Non-Gaussian diffusion}

While in his work Perrin used Einstein's and Smoluchowski's prediction of
the Gaussian PDF (\ref{gauss}) to evaluate his diffusion experiments of an
ensemble of unbiased tracer particles, Kappler employed a torsional balance
setup to measure the angular Brownian motion
of a small mirror to map, with remarkable
accuracy, the resulting Gaussian angle distribution \cite{kappler}. With the
much better experimental resolution of modern experiments, especially those
using single particle tracking \cite{braeuchle}, and simulations, mapping out
the PDF of stochastically or actively moving entities has become routine.
However, more recently a number of systems report significant deviations
from the Gaussian statistic, and such non-Gaussian diffusion has attracted
significant attention from both experimentalists and modellers. After a brief
summary of different systems reporting non-Gaussian PDFs we will introduce
several possible stochastic approaches to explain such non-Gaussianity.

\subsection{Phenomenology}

While one tacitly assumes that the occurrence of the Brownian (or Fickian)
linear scaling $\langle x^2(t)\rangle\simeq K_1t$ in time of the MSD implies
a Gaussian shape of the PDF, intermittent departure from a Gaussian statistic
was reported from disordered solids (glasses, supercooled liquids) \cite{kob,
kob1,vargas} and interfacial dynamics \cite{samanta,skaug}. The ubiquitousness
of "Brownian yet non-Gaussian" (BnG) diffusion was popularised in the field of
soft and biological matter by Granick in their mini-review
\cite{granick}. Two specific systems addressed there are colloidal beads
diffusing along tubes made up of phospholipid bilayer (the simplest building
blocks for biological membranes) and colloidal beads moving in entangled
actin hydrogels. In the actin gel system they observe an exponential shape
("Laplace distribution") of the PDF of the form \cite{granick,granick1}
\begin{equation}
\label{laplace}
P(x,t)=\frac{1}{2\lambda(t)}\exp\left(-\frac{|x|}{\lambda(t)}\right),
\end{equation}
with the decay length $\lambda(t)\simeq t^{1/2}$ scaling like the square root
of time. As can be easily checked this PDF is normalised on the interval
$x\in(-\infty,\infty)$ and indeed encodes the Brownian scaling $\langle
x^2(t)\rangle\simeq\lambda^2(t)\simeq t$. In contrast, in the lipid bilayer
tube system a crossover was observed: for times longer than some characteristic
time the Laplace distribution reverts to a Gaussian shape. In both systems
the Fickian scaling $\langle x^2(t)\rangle\simeq t$ \emph{with a stationary
prefactor\/} is observed at all times.  Similar examples for diffusive motion
yet exponential tails were reported for tracer diffusion in suspensions of
swimming microorganisms \cite{leptos} and colloidal nanoparticles adsorbed
to fluid interfaces \cite{xue,wang,dutta} Similarly, BnG was observed for
the motion of nematodes \cite{hapca}. In a detailed analysis using single
particle tracking of fluorescently labelled colloidal particles in an array of
micropillars, the non-Gaussianity of the resulting ensemble PDF was shown to
be due to an apparent position-dependent, heterogeneous particle diffusivity,
plus a heterogeneous particle distribution \cite{yael}. While the overall
MSD was linear (i.e., Fickian), depending on the micropillar density and
the randomness of their placement in space, the study mapped out the PDFs
of the diffusivities and the apparent anomalous diffusion exponent $\alpha$
for individual particles.  Especially the PDFs for $\alpha$ turn out to be
quite broad. A recent example for BnG dynamics is nanoparticle transport in
a graphene liquid cell \cite{glc}.

Gaussianity is also a hallmark of certain anomalous diffusion processes. The
best known is fractional Brownian motion (FBM), going back to Kolmogorov
\cite{kolmogorov} as well as Mandelbrot and van Ness \cite{mandelbrot}. FBM
is described in terms of an overdamped Langevin equation $\dot{x}(t)=\xi_{
\alpha}(t)$, where $\xi_{\alpha}(t)$ is the driving noise. While Brownian
motion corresponds to a white Gaussian form for $\xi_1(t)$, in FBM the noise
remains Gaussian yet is long-ranged correlated. Specifically, the autocorrelation
of the noise is stationary and of the power-law form $\langle\xi_{\alpha}(t)
\xi_{\alpha}(t+\tau)\rangle=D_{\alpha}\alpha(\alpha-1)\tau^{\alpha-2}$, at long
times \cite{deng,jeon12,pccp}. For subdiffusion ($0<\alpha<1$), the noise-noise
correlator has a negative sign, mirroring so-called antipersistence, while for
superdiffusion ($1<\alpha<2$) a persistent, positively correlated power-law in
$\tau$ is followed. FBM of subdiffusive type is particularly found to
characterise diffusion in viscoelastic systems in the overdamped
limit.\footnote{Strictly, viscoelastic diffusion should be described
in terms of the generalised Langevin equation with power-law memory kernel
\cite{zwanzig,deng,goychuk}, but we here base our discussion on the simpler FBM.}

In a range of systems viscoelastic diffusion with a non-Gaussian PDF was observed.
Thus, single particle tracking in bacteria and yeast cells demonstrated clearly
an exponential distribution of apparent particle diffusivities along with a
Laplace shape of the particle PDF \cite{lampo} (compare also the results in
\cite{stuhrmann,stuhrmann1}). Non-Gaussian diffusion along
with pronounced subdiffusion was also found for the motion of phospholipid
molecules in bilayer membranes at sub-nanosecond times \cite{gupta}. A dynamic
crossover from subdiffusion to normal diffusion and an approximately exponential
displacement PDF was reported for the motion of acetylcholine receptors in the
membranes of living biological cells \cite{he}. Both lipids and membrane proteins
were shown to exhibit pronounced non-Gaussian PDFs in protein-decorated lipid
bilayer membranes at higher crowding fractions \cite{ilpo}: here the PDF was of
a stretched Gaussian form, $P(x,t)\approx\exp(-c|x|^{\kappa})$ with $1<\kappa<2$.
Similar stretched Gaussian shapes were also observed in the (mostly)
superdiffusive, active motion of dictyostelium discoideum amoeba cells spreading
on a surface \cite{carsten}; an interesting observation here is that, within
the experimentally accessible time window, the displacement PDF becomes \emph{
more\/} exponential with \emph{increasing\/} lag time, i.e., the stretching
exponent $\kappa$ tends to unity, contrasting the crossover to a Gaussian in
the colloids-on-nanotube experiment of \cite{granick} or in the graphene
liquid cell experiments in \cite{glc}. A surprisingly rich
dynamic behaviour was found for the lateral diffusion of doxorubicin drug
molecules in between two silica slabs \cite{amanda}: the displacement PDF is
pronouncedly non-Gaussian, but the motion is also antipersistent and non-ergodic.
Crossover from subdiffusion to a plateau of the MSD along with pronounced
non-Gaussianity was reported for tracers in an active DNA gel with FtsK50C
molecular motors \cite{bertrand}.
A rich dynamic behaviour was also reported from glass-forming Lennard-Jones
systems \cite{sandalo}: here the scaling of the length scale $\lambda(t)$ in the
Laplace PDF has the diffusive scaling $\lambda(t)\simeq t^{1/2}$ in the 2D case
but exhibits $\lambda(t)\simeq t^{1/3}$ in 3D and $\lambda(t)\simeq t^{1/4}$ in
4D. A crossover from early subdiffusion to BnG was found in quasi-2D suspensions
of colloidal beads in a spatially random yet static optical force field
\cite{raffaele}.

From simulations of the motion of particles in 2D static disordered landscapes
non-Gaussian behaviour were rationalised in \cite{luo}, and especially the role
of an additional peak in the centre of the displacement PDF discussed \cite{luo1}.
In fact, the precise relaxation dynamics of the central peak in heterogeneous
systems was scrutinised in detail recently \cite{igor_peak}. From extensive
computer simulations the role of the system initiation, i.e., randomly initiated
versus equilibrated particle positions, in BnG was revealed in \cite{zhenya}.
In the context of non-Gaussian diffusion in quenched landscapes we finally note
that random walks visiting traps with exponentially distributed depths, as well
as the annealed continuous time random walk limit with scale-free waiting time
PDFs \emph{always\/} have stretched exponential displacement PDFs \cite{report,
bouchaud,bouchaud_quenched,monthus,stas_quenched,stas_quenched1,scher}.

\subsection{Quantifying non-Gaussianity}

Before turning to concrete models to describe BnG we first mention how BnG can
be quantified. One concrete piece of information already mentioned is the
diffusion length $\lambda(t)$, from which the $x$-$t$ scaling can be inferred.
Non-Gaussianity as such, of course, needs to be identified from higher order
moments. Typically, for a centred process one uses the kurtosis $K=\langle
x^4(t)\rangle/\langle x^2(t)\rangle^2$. For a Gaussian, for instance, $K=3$
in 1D and $K=2$ in 2D. Platykurtic PDFs with $K$ values smaller than that for
a Gaussian, have thinner tails than a Gaussian and are sometimes termed
sub-Gaussian \cite{kahane}. Leptokurtic PDFs, in contrast, have a higher
$K$ value than a Gaussian and thus thicker tails ("super-Gaussians"
\cite{benveniste}), for instance, the Laplace PDF has $K=9$ in 1D and $K=4$
in 2D. An alternative to the kurtosis is the non-Gaussian parameter
\cite{rahman,huang} $(d-2)!!d^2K/(2+d)!!-1$ in $d$ dimensions

A recently introduced measure for non-Gaussianity is the codifference
\cite{jakub1}
\begin{equation}
\label{codiff}
\tau^{\theta}(t)=\frac{1}{\theta^2}\ln\frac{\langle\exp(i\theta[x_{s+t}-x_s])
\rangle}{\langle\exp(i\theta x_{s+t})\rangle\langle\exp(-i\theta x_s)\rangle}
\end{equation}
of a random process $x_t$ with the continuous parameter $\theta$. It can be
argued that the codifference pays more weight to the bulk of the underlying
PDF, in contrast to the covariance. For Gaussian variables with the variance
$\sigma^2$ it has the simple form $\exp(-[\theta\sigma^2]/2)$ and has the
characteristic behaviour of a memory function. From detailed analysis it can
be shown that the codifference (\ref{codiff}) is in fact a suitable measure
to detect non-Gaussianity (and non-ergodic behaviour) \cite{jakub1}. We also
mention the availability of methods based on random coefficient autoregressive
approaches, that can be mapped on BnG-style models (see below) \cite{jakub2}.

\subsection{Superstatistical approaches}

Granick and coworkers proposed that the non-Gaussian displacement statistic
$P(x,t)$ can be understood in terms of the superposition
\begin{equation}
\label{suppdf}
P(x,t)=\int_0^{\infty}G(x,t|D)p(D)dD,
\end{equation}
where $G(x,t|D)$ represents a Gaussian of the form (\ref{gauss}) for a specific
value $D$ of the diffusion coefficient, and $p(D)$ is a PDF of $D$ values. In
this view each particle performs normal diffusion but with a different $D$.
This scenario could simply represent particles of different sizes or particles
moving in patches of different local properties.\footnote{In the latter case the
formulations only holds as long as the particles do not cross boundaries between
patches with other $D$ values---in contrast to the annealed transit time model
\cite{massignan}.} The MSD of this process reads
\begin{equation}
\label{supmsd}
\langle x^2(t)\rangle=2t\int_0^{\infty}Dp(D)dD=2\langle D\rangle t,
\end{equation}
that is, the MSD is linear with the effective diffusion coefficient $\langle
D\rangle$. The approach (\ref{suppdf}) is in fact the "superstatistics"
approach by Beck and Cohen \cite{beck}, in the sense that the statistical
behaviour of the Gaussian PDF $G(x,t|D)$ is modulated by additional averaging
over the diffusivity PDF $p(D)$. In this superstatistical approach one can
show that the Laplace PDF (\ref{laplace}) corresponds uniquely to an
exponential form of $p(D)$ \cite{chechkin_prx}. In \cite{hapca} a gamma
distribution for $p(D)$ was used, compare also \cite{vittoria}. Power-law
forms of $p(D)$ lead to superstatistical forms of the PDF (\ref{suppdf})
with power-law tails \cite{chechkin_prx,jain}, another relevant form for
a superstatistical PDF are stretched Gaussians \cite{news,chechkin_prx}.
Superstatistical approaches were also formulated for non-Fickian,
anomalous diffusion dynamics, especially extensions of the generalised
Langevin equation with random parameters \cite{vdstraeten,jakub}. In particular,
in \cite{jakub} it is shown that this formulation may also lead to more
exotic shapes for the PDF, such as a Cauchy law.

Historically, concepts similar to superstatistics were used before, notably,
in the field of turbulence, where the refined similarity hypotheses take
into account fluctuations of energy dissipation and lead to intermittent
corrections to the famous -5/3 spectrum of energy in the inertial interval.
In this setting
non-Gaussian cascades were considered to emerge from "statistical mixing"
by Obukhov and Kolmogorov \cite{obukhov,obukhov1,obukhov2}. This idea was
further developed by Castaing and coworkers \cite{castaing,naert} and is
related to the scale dependence of velocity increments in turbulent systems
\cite{friedrich,marcq}. Other fields using concepts similar to superstatistics
include financial mathematics ("compounding") \cite{dubey} and the study of
fracture processes \cite{vallia}.

A process closely related to superstatistics is generalised grey Brownian
motion (ggBM), defined in terms of the stochastic equation \cite{lumapa:ggBM,
main1:ggBM,main2:ggBM,Mura:ggBM,daniel}
\begin{equation}
\label{ggBM_x}
x_\mathrm{ggBM}(t)=\sqrt{2D}B(t)
\end{equation}
for the particle trajectory $x_\mathrm{ggBM}(t)$, see also the discussions
in \cite{vittoria,jakub1}. $D$ here is now a random diffusivity. The core idea
is that different yet physically identical particles move in disjointed areas
in which their diffusivity is different---in other words,, the essential view
of superstatistics. We could also think of physically different particles,
each with a different diffusivity, in a homogeneous environment. GgBM includes
anomalous diffusion scenarios, as detailed further in \cite{main1:ggBM,
Mura:ggBM}.

\subsection{Diffusing-diffusivity models}

In typical (anomalous) diffusion models we assume that a particle is
characterised by a given diffusion coefficient or noise strength. This includes
the superstatistical approach, in which each particle has its own diffusion
coefficient, that does not change in time. This is not always justified. For
instance, lipid molecules in crowded bilayer membranes are found to exhibit
intermittent dynamics switching between high and low diffusivity modes, and
this behaviour can be mimicked by hard core particles moving through a fixed
obstacle environment \cite{ilpo}. Another scenario is that a particle is moving
in a hydrogel with (sufficiently fast) local fluctuations of mesh sizes, such
that on its way the particle experiences a varying degree of obstacle densities
\cite{yann}.

Another scenario emerges for certain proteins moving freely in aqueous solution.
These protein molecules can then be shown to exhibit continuous changes between
conformations, some of which are compact while others are extended. These
conformations correspond to different local minima of the free energy
landscape of the protein. Significant temporal fluctuations of the effective
size (measured in terms of the gyration radius $R_g$) and thus the hydrodynamic
radius of such a perpetually shape-shifting protein molecule were demonstrated
to effect a stochastic instantaneous diffusivity. This instantaneous
diffusivity fulfils a Stokes-Einstein-type relation with $R_g$ \cite{eiji}.
As further detailed in section \ref{poly} tracer particles may also exhibit
a stochastically changing diffusivity due to ongoing (de)polymerisation
size-variations.

The motion of a tracer particle with a stochastically evolving diffusivity
can be modelled in terms of a Langevin equation in the so-called "diffusing
diffusivity" approach \cite{gary}. This model was further developed in
\cite{jain,jain1,yann,tyagi,grebenkov}. In a minimal formulation the
diffusing-diffusivity model is captured by the set of coupled stochastic
equations \cite{chechkin_prx}\footnote{In \cite{chechkin_prx} the model is
considered in arbitrary dimensions for position $x$ and auxiliary variable
$y$.}
\begin{subequations}
\label{langmin}
\begin{eqnarray}
\label{lang1}
\frac{d}{dt}x(t)&=&\sqrt{2D(t)}\xi(t),\\
\label{lang2}
D(t)&=&y^2(t),\\
\label{lang3}
\frac{d}{dt}y(t)&=&-\frac{1}{\tau}y+\sigma\eta(t).
\end{eqnarray}
\end{subequations}
Equation (\ref{lang1}) here represents the Langevin equation for the particle
position $x(t)$, driven by the white Gaussian noise $\xi(t)$. In contrast to
the standard Langevin equation, however, the diffusion coefficient $D(t)$
is explicitly time-dependent. The dynamics of this noise strength $D(t)$ is
specified by equations (\ref{lang2}) and (\ref{lang3}). First, to vouchsafe
positivity of the diffusivity, we write $D(t)$ as the square of the auxiliary
variable $y(t)$. The dynamics of $y(t)$ in expression (\ref{lang3}) is that
of an Ornstein-Uhlenbeck process \cite{vankampen}, that is, diffusion driven
by the white Gaussian noise $\eta(t)$ in the presence of a Hookean restoring
force. The motion $y(t)$ is thus confined by an harmonic potential and will
reach equilibrium beyond the correlation time $\tau$. The physical dimension
of the noise strength in the Ornstein-Uhlenbeck process for the
auxiliary variable $y(t)$ here is $[\sigma]=\mathrm{cm}/\mathrm{sec}$.

The diffusing-diffusivity dynamics encoded in equations (\ref{langmin}) can
be shown to reproduce the superstatistical exponential tails (with dimension
dependent power-law correction) at times shorter than $\tau$,
\begin{equation}
P(x,t)\sim\frac{1}{\sqrt{2\pi|x|\sigma(\tau t)^{1/2}}}\exp\left(
-\frac{|x|}{\sigma(\tau t^{1/2})}\right).
\end{equation}
At times $t\gg\tau$, when correlations in the diffusivity process $D(t)$ (or,
better, in the auxiliary variable $y(t)$) are relaxed, the Gaussian
\begin{equation}
P(x,t)\sim\frac{1}{\sqrt{4\pi\langle D\rangle t}}\exp\left(-\frac{x^2}{4\langle
D\rangle t}\right)
\end{equation}
is recovered. Concurrently to the crossover from Laplace-type PDF to a Gaussian
shape, the MSD of the minimal diffusing-diffusivity process (\ref{langmin}) is
given by the law
\begin{equation}
\langle x(t)\rangle=2\langle D\rangle t,\quad \langle D\rangle=\frac{1}{2}
\sigma^2\tau
\end{equation}
at all times, as long as an equilibrium initial condition for $y(t)$ is chosen.
Moreover, the initial non-Gaussianity can be monitored in terms of the kurtosis
$K=\langle x^4(t)\rangle/\langle x^2(t)\rangle^2$, that shows a crossover from
the short-time value $K\sim9$, the value for an exponential PDF, to $K\sim3$,
the value for a Gaussian, in the 1D case. All these results were also shown to
be in full agreement with stochastic simulations \cite{chechkin_prx}.

We note that more technically, the minimal diffusing-diffusivity model
(\ref{langmin}) can be shown to correspond to a dynamic subordinated to a
Brownian diffusion process, and it can be formulated in terms of a bivariate
Fokker-Planck equation \cite{chechkin_prx}. In \cite{vittoria} the model was
extended to a generalised gamma distribution in the superstatistical,
short-time limit, and the initial condition for $y(t)$ was generalised to
non-equilibrium initial conditions, leading to distinct crossover behaviours
in the MSD. We finally note that diffusing-diffusivity models analogous to
the minimal model exist, in different language, in mathematical finance.
Notably, the Heston model \cite{heston} with its stochastic volatility.
In turn the Heston model is a specific case of the Cox-Ingersoll-Ross
model \cite{cox}. We also note that the crossover
behaviour governed by equations (\ref{langmin}) with the stationary
dynamics for the stochastic diffusion coefficient $D(t)$ is generically
different from formulations with diffusion coefficients $D(t)$ based on
deterministic dynamics such as scaled Brownian motion, effective FBM
models in absence of boundaries, or combinations based on such
processes \cite{lutz,jae_sbm,anna,igor_sbm,paolo_sbm,paolo_sbm1}.

The generic behaviour encoded in the minimal diffusing-diffusivity approach
of equations (\ref{langmin}) is quite robust and matches (up to minor
details) the short- and long-time behaviour of the other
diffusing-diffusivity models developed in
\cite{gary,yann,grebenkov,jain,jain1,tyagi}. What happens when we combine
the diffusing-diffusivity idea with Gaussian anomalous diffusion of the FBM
type? This was analysed in \cite{wei} where the Ornstein-Uhlenbeck dynamic
(\ref{lang3}) for the time-dependent diffusivity $D(t)$ in (\ref{lang2})
was combined with a Langevin equation $\dot{x}(t)=\sqrt{2D(t)}\xi_{\alpha}
(t)$ driven by fractional Gaussian noise chosen in the Mandelbrot-van Ness
smoothed form $\langle\xi_{\alpha}(t)\xi_{\alpha}(t+
\tau)=(2\delta^2)^{-1}(|\tau+\delta|^{\alpha}-2|\tau|^{\alpha}+|\tau-
\delta|^{\alpha})$ for the noise autocorrelation and small constant
$\delta$. For antipersistent noise $\xi_{\alpha}$ with
$0<\alpha<1$ a crossover from initial subdiffusive scaling of the MSD to
normal-diffusive scaling was obtained. In contrast, for persistent noise
($1<\alpha<2$) both short- and long-time scaling of the MSD was found to
be superdiffusive. As desired, the PDF crosses over from a Laplace-type
distribution at short times to a long-time Gaussian. The effective
diffusivity as function of the anomalous diffusion exponent exhibits a
discontinuity around $\alpha=1$. This behaviour was demonstrated to be
different from the generalised model of \cite{tyagi} and the two-state
model developed in \cite{diego}. Thus, the interplay of the long-ranged
noise correlations and the crossover dynamics of the diffusion coefficient
effect a quite intricate, non-universal dynamics. In fact, the often
unexpected behaviour of FBM was recently also highlighted in its behaviour
in the presence of reflecting boundaries with possible connection to
brain fibre density fields \cite{vojta,tobias,vojta1,vojta2,
vojta3,tobias1,skirmantas}.

We finally mention that diffusing-diffusivity dynamics in association with
active Brownian motion was recently analysed in \cite{sabine}, finding
a crossover from a short-time exponential shape of the PDF to a long-time
Gaussian form.

\subsection{Large-deviation approach}

While we mentioned observations of non-Gaussian statistics in the form of
stretched Gaussian shapes, the majority of cases reported involve Laplace
PDFs with their characteristic exponential tails. In terms of a random
walk approach the emergence of such unconventional tails can be understood
by extreme-value arguments, as demonstrated in \cite{stas,stas1} for general
continuous-time random walk processes. The central argument goes as follows:
when we derive the Gaussian limit PDF using the law of large numbers or the
central limit theorem, the main condition is the limit of many jumps, that
is, the long-time limit. What is the shape of the tails when the number of
jumps or time is finite? In fact, the tails of the PDF $P(x,t)$ are then
dominated by extreme events. Applying the theorems of large-deviation
statistics, the leading contribution to the PDF in the limit $|x|/t\to
\infty$ is given by \cite{stas}
\begin{equation}
\label{largedev}
P(x,t)\sim\exp\left(-\left[\kappa\log\left(\frac{|x|}{t}\right)^{1-1/\beta}
\frac{|x|}{t}+C\right]t\right).
\end{equation}
i.e., leading exponential tails with power-law corrections. The authors
state that "the exponential tails for the positional PDF are rather a
rule and not an exception. The exponential decay of the tails is a
general feature exactly like the Gaussian behaviour (that is dictated
by the CLT [central limit theorem]) at the centre." Note the somewhat
different form of the large-deviation result (\ref{largedev}) compared to
the BnG form (\ref{laplace}): in (\ref{largedev}) no scaling between $x$
and $t$ occurs, in analogy to the results reported in \cite{kob1}.

\subsection{Polymerisation models}
\label{poly}

In the above example of the perpetually shape-shifting protein molecule
\cite{eiji} the stochastic diffusivity dynamics was effected by the change
of the effective protein size as measured by the gyration radius. A similar
effect can be observed when the tracer particle is growing or shrinking.
In fact, \emph{ARG3} messenger RNA molecules were reported to occur with a
broad distribution of diffusivities, associated, inter alia, with
conglomeration with one or multiple mRNA and other associated entities such
as RNA-binding proteins \cite{moerner}.

In a simple polymerisation dynamics approach $A_N+A_1\stackrel[k_-]{k_+}{
\rightleftharpoons}A_{N+1}$ describing the addition of a monomer $A_1$ to a
polymer $A_N$ containing $N$ monomers with rate $k_+$ and the backwards
reaction, chains with a time-dependent, fluctuating polymerisation degree
are being formed \cite{fulvio}. If a monomer has diffusivity $D_0$ then a
Stokes-Einstein-type relation emerges in the form $D_N\simeq D_0/N^{\nu}$,
where the scaling exponent depends on the specific polymer model. Thus,
for the Rouse limit (Gaussian chain), $\nu=1/2$, $\nu=1$ for the Zimm
model when hydrodynamic interactions are included, and $\nu=2$ in the
reptation model when the tagged polymer moves in a solution of entangled
polymers \cite{doi}. Detailed analysis and simulations then demonstrate
that the resulting diffusion of such a polymer with fluctuating size is
non-Gaussian at short times and crosses over to a Gaussian diffusion at
long times, corresponding to the equilibrium size distribution of the
polymer chain. It is shown in \cite{fulvio} that the kurtosis in the
non-Gaussian state can assume extremely high values. A complementary
analysis based on simulations and analytical arguments is presented in
\cite{mario}.

\subsection{Random-diffusivity models}

The concept of diffusing-diffusivity is further carried on in \cite{sposini}
based on a Langevin equation of the form $\dot{x}(t)=\sqrt{2D_0\Psi_t}\xi(t)$,
where $\xi$ is white Gaussian noise and $\Psi_t$ a positive-definite random
function of time. A squared Ornstein-Uhlenbeck process for $\Psi_t$ leads
back to the minimal diffusing-diffusivity model (\ref{langmin}), while other
choices discussed in \cite{sposini} are jump processes with Gamma and
L{\'e}vy-Smirnov distributions, as well as the cases of squared Brownian motion
($\Psi_t=B(t)^2$), rectified Brownian motion ($\Psi_t=\Theta(B(t))$, where
$\Theta(\cdot)$ is the Heaviside step function), and geometric Brownian
motion ($\Psi_t=\exp(-B(t)/a)$). The emerging non-Gaussian shapes for the
displacement PDF and the corresponding single trajectory power spectra (see
\cite{gleb,gleb1}) are calculated for each case.

\subsection{Random-coefficient autoregressive models}

The Langevin equation approach is a standard approach in non-equilibrium
statistical physics. In other fields such as financial mathematics, so-called
autoregressive models are widespread \cite{armodels}. The connection between
both can be directly established in certain cases. It can then be shown that
if we start from the generalised Langevin equation $dv(t)=-\Lambda(t)v(t)+
\sqrt{D(t)}dB(t)$ with the time-dependent friction and diffusion coefficients
$\Lambda(t)$ and $D(t)$, and where $B(t)$ represents Brownian motion (Wiener
process), this dynamics can be mapped onto a random-coefficient autoregressive
model (rcAR). Namely, discretisation using $v_k=v(k\Delta t)$ with small but
finite time delay $\Delta t$ and integer $k$, leads to the approximation
\begin{equation}
\label{rcar}
v_k-(1-\Lambda_k\Delta t)v_{k-1}=\sqrt{D_k}\Delta B_k.
\end{equation}
This is indeed an autoregressive process of the class rcAR(1) with AR
coefficient $1-\Lambda_k\Delta t$ \cite{jakub2}. Based on such mappings one
can then use the well-established methods of data analysis in autoregressive
modelling to identify measured data exhibiting BnG \cite{jakub2}.

\subsection{Mobile-immobile models}

We finally mention a simple model describing relaxation in a system with two
populations of particles, mobile and immobile ones, that mimics some features
of BnG. As was demonstrated in \cite{mora} Brownian diffusion in a dilute field
of traps is Fickian but non-Gaussian. Briefly, the starting point are the
coupled equations for the mobile ($n_m$) and immobile ($n_{im}$) particle
concentrations,
\begin{subequations}
\label{mim}
\begin{eqnarray}
\frac{d}{dt}n_m(\mathbf{r},t)=-\beta n_m(\mathbf{r},t)+\frac{1}{\tau}
n_{im}(\mathbf{r},t)+D\nabla^2n_m(\mathbf{r},t),\\
\frac{d}{dt}n_{im}(\mathbf{r},t)=\beta n_m(\mathbf{r},t)-\frac{1}{\tau}
n_{im}(\mathbf{r},t)
\end{eqnarray}
\end{subequations}
where $\mathbf{r}(t)=(x(t),y(t))$ is a two-dimensional vector. The initial
conditions are $n_m(\mathbf{r},0)=N_0\delta(\mathbf{r})$ and $n_{im}(\mathbf{
r},0)$. Note than in \cite{mora} equilibrium initial conditions were used.
Such a choice may be more natural for some systems, however, this choice
weakens the net effect. For that reason we here choose the non-equilibrium
condition that all particles initially are mobile.

The Laplace transform $P(\mathbf{r},s)=\mathscr{L}\{P(\mathbf{r},t)\}=
\int_0^{\infty}P(\mathbf{r},t)\exp(-st)dt$ of the PDF is obtained after
Fourier-Laplace transformation of equations (\ref{mim}) in terms of the
modified Bessel function $K_0$ as
\begin{equation}
\label{morapdf}
P(\mathbf{r},t)=\frac{1}{N_0}\Big(n_m(\mathbf{r},s)+n_{im}(\mathbf{r},s)
\Big)=\frac{1}{2\pi D}\frac{\varphi(s)}{s}K_0\left(r\sqrt{\frac{\varphi(s)}
{D}}\right),
\end{equation}
where
\begin{equation}
\varphi(s)=\frac{s}{s+1/\tau}\left(s+\beta+\frac{1}{\tau}\right).
\end{equation}
The MSD encoded in this PDF reads
\begin{equation}
\label{moramsd}
\langle\mathbf{r}^2(t)\rangle=\frac{4D}{1+\beta\tau}\left(t+\frac{\beta
\tau^2}{1+\beta\tau}\left[1-\exp\left(-\frac{1+\beta\tau}{\tau}t\right)
\right]\right).
\end{equation}
As can be seen from equations (\ref{morapdf}) and (\ref{moramsd}), for
$\beta\tau\gg1$ the MSD grows linearly at short and long times while the
PDF is Gaussian. However, at intermediate times $\beta^{-1}\ll t\ll\tau$
the MSD exhibits a plateau-like behaviour while the PDF has the
time-independent shape
\begin{equation}
P(\mathbf{r},t)\approx P(\mathbf{r})=\frac{\beta}{2\pi D}K_0\left(\sqrt{
\frac{\beta}{D}}r\right)\simeq\exp\left(-\sqrt{\frac{\beta}{D}}r\right),
\end{equation}
with exponential tails. Thus, the crossover from the plateau- to normally
diffusive-regime in the MSD is accompanied by a crossover from exponential
to Gaussian behaviour in the shape of the PDF.

\section{Conclusions}

Non-Gaussianity has been observed in a large number of systems in many
different systems. Similar to the occurrence of non-Brownian, anomalous
diffusion, non-Gaussianity breaks with one of the dogmas of conventional
statistical physics, the predominance of the central limit theorem. The MSD
in these processes can be Fickian or exhibit anomalous scaling, while the
displacement PDFs can assume a Laplace PDF with exponential tails, stretched
Gaussian shapes, or even power-laws. In a given experiment or simulation,
the non-Gaussian character is either preserved within the accessible
experimental window of probed time scales, or crossover behaviours can be
observed. The latter typically show the emergence of a Gaussian PDF beyond
some correlation time, although other observations exist. A growing number of
stochastic processes are being devised to describe this non-Gaussian behaviour.

The approaches reviewed here correspond to \emph{annealed\/} models, in which
the instantaneous value of the particle diffusivity is unconnected from the
particle's specific position in space. This is justified when the particles
of an ensemble themselves have a distribution of diffusivities, and then the
superstatistical description with its time-independent diffusivity PDF serves
as the apt physical description. Other scenarios for annealed descriptions
are those of shape-shifting or polymerising tracers, or when the environment
changes rapidly enough such that the particle experiences a renewed value of
its local mobility compared to its previous visit to that position. In three
dimensions the annealed models provide reasonable approximations even if the
environment does not change rapidly, as here the probability to revisit
specific positions is relatively low. Models with explicit quenched disorder,
when correlations in the particle motion build up while the particle samples
the same mobility values each time it returns to previously visited points,
investigating the emerging non-Gaussianity of the diffusive spreading
include predominantly simulations-based information \cite{luo,luo1,zhenya,
sandalo,jakubstas}. Finding analytical access to the quantitative modelling
in such cases remains a challenge. Equally elusive is the understanding of
the peculiar dimension dependence of the scaling $\lambda(t)\simeq t^{1/d}$
with spatial dimension $d=2,3,4$ in the Lennard-Jones system in \cite{sandalo}.

\ack

RM acknowledges the German Science Foundation (DFG, grant no. ME 1535/12-1) for
support. RM also acknowledges the Foundation for Polish Science (Fundacja na
rzecz Nauki Polskiej, FNP) for support within an Alexander von Humboldt
Honorary Polish Research Scholarship. AC acknowledges support of the Polish
National Agency for Academic Exchange (NAWA).

\end{document}